\documentclass[conf]{new-aiaa}
\usepackage[utf8]{inputenc}

\usepackage{graphicx}
\usepackage{amsmath}
\usepackage[version=4]{mhchem}
\usepackage{siunitx}
\usepackage{longtable,tabularx}
\usepackage{caption}
\usepackage{multirow}
\usepackage{subcaption}
\usepackage{ulem}
\usepackage{xcolor}

\title{Features of a Splashing Drop on a Solid Surface and \\the Temporal Evolution extracted through \\Image-Sequence Classification using \\an Interpretable Feedforward Neural Network}

\author{Jingzu Yee\footnote{PhD Candidate, Department of Mechanical Systems Engineering.}, 
Daichi Igarashi\footnote{Master Student, Department of Mechanical Systems Engineering}, 
Akinori Yamanaka\footnote{Professor, Department of Mechanical Systems Engineering.}, and 
Yoshiyuki Tagawa
\footnote{Professor, Department of Mechanical Systems Engineering}
\footnote{Professor, Institute of Global Innovation Research}
}
\affil{Tokyo University of Agriculture and Technology, 2-24-16, Naka-cho, Koganei, Tokyo, 184-8588}

\begin{document}

\maketitle

\begin{abstract}
This paper reports the features of a splashing drop on a solid surface and the temporal evolution, which are extracted through image-sequence classification using a highly interpretable feedforward neural network (FNN) with zero hidden layer.
The image sequences used for training-validation and testing of the FNN show the early-stage deformation of milli-sized ethanol drops that impact a hydrophilic glass substrate with the Weber number ranges between 31--474 (splashing threshold about 173).
Specific videographing conditions and digital image processing are performed to ensure the high similarity among the image sequences.
As a result, the trained FNNs achieved a test accuracy higher than 96\%.
Remarkably, the feature extraction shows that the trained FNN identifies the temporal evolution of the ejected secondary droplets around the aerodynamically lifted lamella and the relatively high contour of the main body as the features of a splashing drop, while the relatively short and thick lamella as the feature of a nonsplashing drop.
The physical interpretation for these features and their respective temporal evolution have been identified except for the difference in contour height of the main body between splashing and nonsplashing drops.
The observation reported in this study is important for the development of a data-driven simulation for modeling the deformation of a splashing drop during the impact on a solid surface.
\end{abstract}

\section{Nomenclature}

{\renewcommand\arraystretch{1.0}
\noindent\begin{longtable*}{@{}l @{\quad=\quad} l@{}}
$\mathbf{b}$ & bias vector\\
$C$ & number of output classes of classification\\
$H$ & impact height\\
$h_\mathrm{img}$ & image height in terms of pixel\\
$l$ & binary cross-entropy loss\\
$M$ & total number of pixels in an image sequence\\
$N_\mathrm{img}$ & total number of images in an image sequence\\
$\mathbf{q}_\mathrm{out}$ & output vector of the linear function that connects input and output layers of feedforward neural network (FNN)\\
$R_0$ & area-equivalent radius\\
$\mathbf{s}_\mathrm{in}$ & vector of an image sequence\\
$U_0$ & impact velocity\\
$\mathbf{W}$ & weight matrix\\
$\mathbf{w}_1$ & weight vector for computing the probability of an image sequence for being a nonsplashing drop\\
$\mathbf{w}_2$ & weight vector for computing the probability of an image sequence for being a splashing drop\\
$w_\mathrm{img}$ & image width in terms of pixel\\
$We$ & Weber number\\
$\mathbf{y}_\mathrm{pred}$ & probabilities vector for the classification of an image sequence\\
$\mathbf{y}_\mathrm{true}$ & vector of the true label of an image sequence\\
$z_0$ & central height of an impacting drop\\

$\gamma$ & surface tension\\
$\mu$ & dynamic viscosity\\
$\rho$ & density\\
$\sigma$ & sigmoid function\\

\end{longtable*}}

\section{Introduction}
\lettrine{T}{he} impact of a liquid drop on a solid surface is an important high-speed phenomenon to aviation field that can be found in technical applications such as spray cooling, gas turbine fuel injection, aircraft icing, among others \cite{burzynski2020splashing,canders2019cooling,zhang2016effect,zhao2018splashing}.
With recent technology of high-speed videography, the observation of micro-sized drop impact at the time scale of nanoseconds is now possible \cite{thoroddsen2008high, visser2015dynamics, visser2012microdroplet}.
Notably, by using a high-speed camera, Lagubeau \textit{et al.} \cite{lagubeau2012spreading} have successfully observed the spreading dynamics, i.e., the evolution of the drop shape during the impact, which was previously described theoretically and numerically \cite{eggers2010drop, roisman2009inertia}.
Nevertheless, owing to the multiphase nature of this phenomenon (that involves the liquid drop, the solid surface, and the ambient air), many important but nonintuitive characteristics could be missed when observation is carried out with the naked eye alone \cite{josserand2016drop, yarin2006drop, rioboo2001outcomes, kim2014drop, yokoyama2022droplet, usawa2021large, hatakenaka2019magic}.
Especially, the observation gets further complicated when splashing occurs, i.e., the impacting drop breaks up and ejects secondary droplets \cite{gordillo2019note,riboux2014experiments,riboux2017boundary,yokoyama2022droplet, usawa2021large, hatakenaka2019magic, kim2014drop}, instead of just spreading over the surface until it reaches the maximum radius \cite{gordillo2019theory,clanet2004maximal}.

To address this issue, an artificial neural network (ANN), which is a supervised machine learning algorithm inspired by biological neural networks \cite{rosenblatt1958perceptron, hornik1989multilayer, hornik1991approximation, chen2019design}, was proposed to aid the observation of the phenomenon of drop impact in our previous study \cite{yee2022image}.
Although ANNs have been widely utilized and have proven effective in carrying out computer vision tasks such as classification and prediction based on images or videos \cite{krizhevsky2012imagenet, he2016deep,voulodimos2018deep}, they usually function as black boxes.
In other words, the underlying reasoning that leads an ANN to a specific decision is often unknown or not properly understood \cite{arrieta2020explainable, adadi2018peeking}.
By solving the issue of explainability and interpretability, ANNs can become powerful tools for advancing the knowledge towards physical phenomena.

In our previous study, a highly interpretable ANN model, namely a feedforward neural network (FNN) with zero hidden layer, was trained and achieved an accuracy higher than 90\% in classifying splashing and nonsplashing drops based on images that show the instantaneous shape of the impacting drops.
This was made possible by the use of highly similar images that were collected under carefully designed videographing conditions and processed with specific digital image processing.
Subsequently, the image features that the FNN identified for classification could be extracted through analyzing the image classification process.
An interesting finding from the study is that the trained FNN identifies the contour height of the main body of the impacting drop as one of the important features for distinguishing splashing and nonsplashing drops.
Although several aspects of drop impact were analyzed and discussed with the aim to identify the possible mechanism underlying the difference in contour height between splashing and nonsplashing drops, it is still unclear at the point of writing.

To further investigate on the reported features and to observe the drop deformation during the impact, the focus of this paper is on the videos or image sequences that shows the temporal evolution of splashing and nonsplashing drops during the early stage of impact.
Remarkably, owing to the high-similarity of the image data, video or image-sequence classification could be performed at a high accuracy by using a FNN with the architecture similar to that used for image classification in our previous study.
The computational process of the well-trained FNN is then visualized to extract the features.
Finally, the physical interpretation and the temporal evolution of the extracted features are analyzed and discussed.

The details are reported in this paper, which is structured as follows.
The methodology of the study, which includes the description of the dataset and implementation of the FNN, is explained in Sec.~\ref{sec:method}.
As for the results, the classification performance of the trained FNN, the process of features extraction, and the physical interpretation and the temporal evolution of the extracted features are analyzed and discussed in Sec.~\ref{sec:result}.
Last, but not least, the concluding remarks and the future prospects of this study are presented in Sec.~\ref{sec:conclusion}.


\section{\label{sec:method}Methodology}
In this section, the dataset of image sequences that show the temporal evolution of the drops during the early stage of impact (Section~\ref{sec:dataset}) and the implementation of the feedforward neural networks (FNNs) for image-sequence classification (Section~\ref{sec:ann}) are explained.

\subsection{\label{sec:dataset} Dataset and Fivefold Cross Validation}
\begin{figure}
\centering
\includegraphics[width=0.45\textwidth]{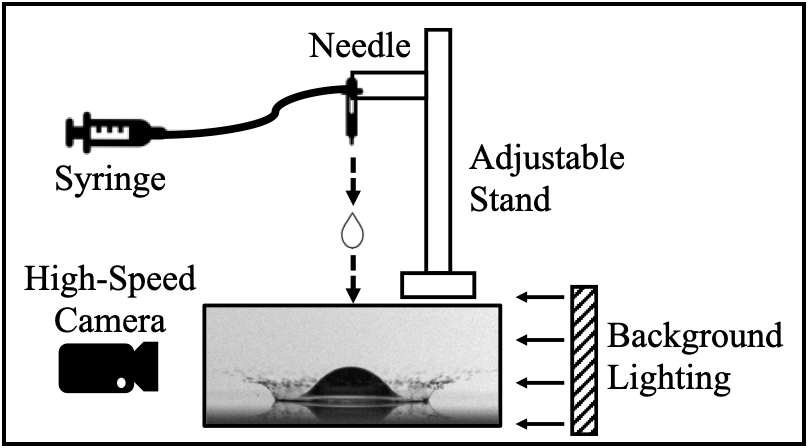}
\caption{\label{fig:setup} Schematic of the experimental setup used during our previous study \cite{yee2022image} to collect the image data.}
\end{figure}
\begin{table}
\centering
\caption{
\label{tab:data_num} 
Numbers of splashing and nonsplashing data for training-validation and testing in each data combination.
The numbers of splashing and nonsplashing data for training-validation or testing are similar among the data combinations.
}
\begin{tabular}{|c|c|c|c|c|c|c|c|}
\hline
\multirow{3}{*}{Combination}
& \multicolumn{7}{c|}{Number of Data} \\\cline{2-8}
& \multicolumn{3}{c|}{Training-Validation} & \multicolumn{3}{c|}{Testing} & Total \\\cline{2-7}
& Splashing& Nonsplashing& Total& Splashing& Nonsplashing& Total& \\\hline
1& 114& 87& 201& 27& 21& 48& 249\\ \hline
2& 112& 86& 198& 29& 22& 51& 249\\ \hline
3& 113& 85& 198& 28& 23& 51& 249\\ \hline
4& 114& 85& 199& 27& 23& 50& 249\\ \hline
5& 111& 89& 200& 30& 19& 49& 249\\ \hline
\end{tabular}
\end{table}

The image data were collected during our previous study \cite{yee2022image}, by using the experiment setup shown in Fig.~\ref{fig:setup}.
They are image data of an ethanol drop (Hayashi Pure Chemical Ind., Ltd.; density $\rho = 789~\rm{kg\:m^{-3}}$, surface tension $\gamma = 2.2 \times 10^{-2}~\mathrm{N\:m^{-1}}$, and dynamic viscosity $\mu = 10^{-3}$~Pa\:s) impacting on a hydrophilic glass substrate (Muto Pure Chemicals Co., Ltd., star frost slide glass 511611) after free-falling from impact height $H$ ranging from $4$--$60$~cm.
The area-equivalent radius of the drop in each of the image data, which is measured before impact, is $R_{0} = (1.29 \pm 0.04)\times 10^{-3}$~m.
The impact velocity $U_0$ and Weber number $We$ $(= \rho U_{0}^2 R_{0}/\gamma)$ range between $0.82$--$3.18~\mathrm{m\:s^{-1}}$ and $31$--$474$, respectively.
The splashing threshold in terms of impact height and Weber number are $H = 20~\mathrm{cm}$ and $We = 173$, respectively.
There are a total of 249 image data: 141 of splashing drops and 108 of nonsplashing drops.
Each of the image data is labeled according to the outcome: splashing or nonsplashing.

\begin{figure}
\centering
\subfloat[]{
\includegraphics[width=0.6\textwidth]{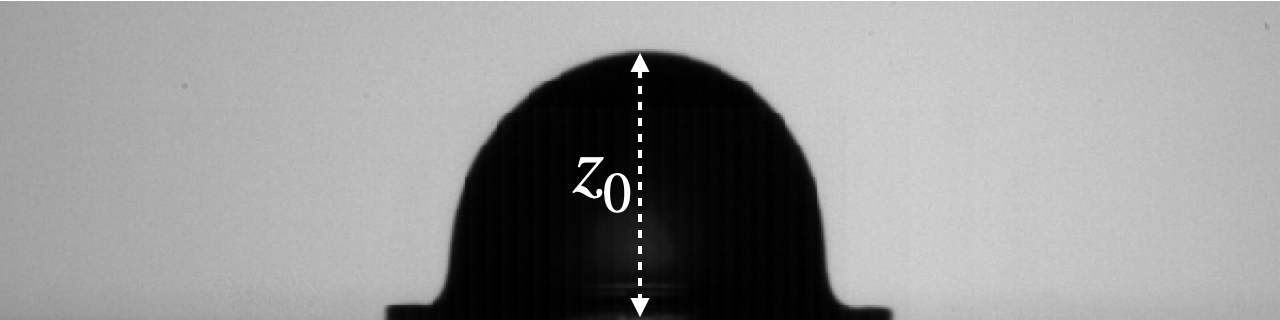}
\label{fig:z0_2R0_075}}
\hfill
\subfloat[]{
\includegraphics[width=0.6\textwidth]{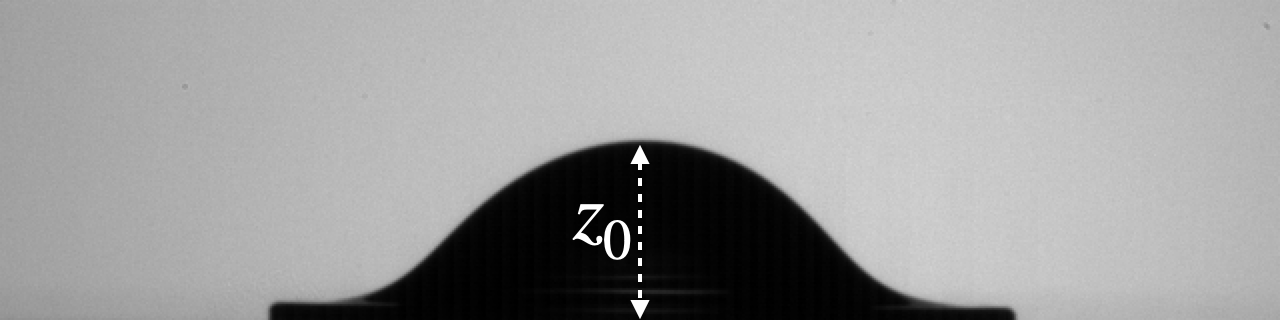}
\label{fig:z0_2R0_050}}
\hfill
\subfloat[]{
\includegraphics[width=0.6\textwidth]{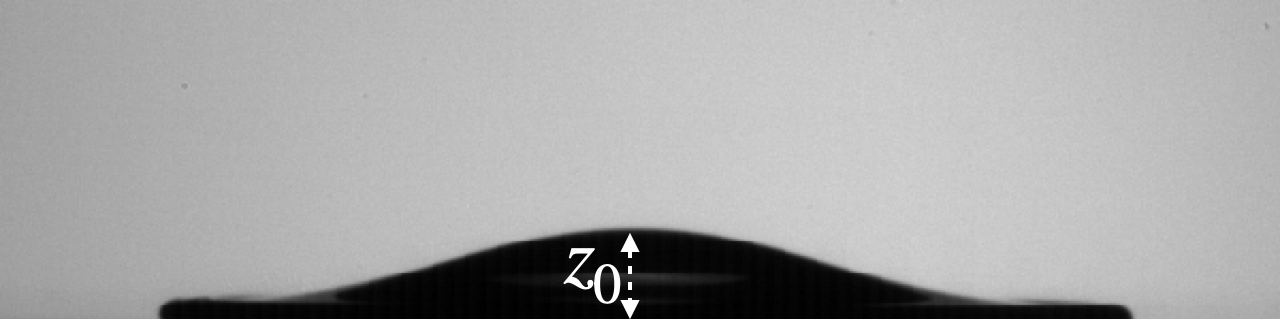}
\label{fig:z0_2R0_025}}
\caption{\label{fig:z0_2R0} 
Illustration of the central height $z_0$ (white-dashed line) of a nonsplashing drop impact with impact height $H = 8~\mathrm{cm}$ and Weber number $We = 73$ when the normalized central height $z_0/2R_0 =$ (\protect\subref*{fig:z0_2R0_075}) $\mathrm{0.75}$, (\protect\subref*{fig:z0_2R0_050}) $\mathrm{0.50}$, and (\protect\subref*{fig:z0_2R0_025}) $\mathrm{0.25}$.
}
\end{figure}

\begin{figure}
\centering
\includegraphics[width=\textwidth]{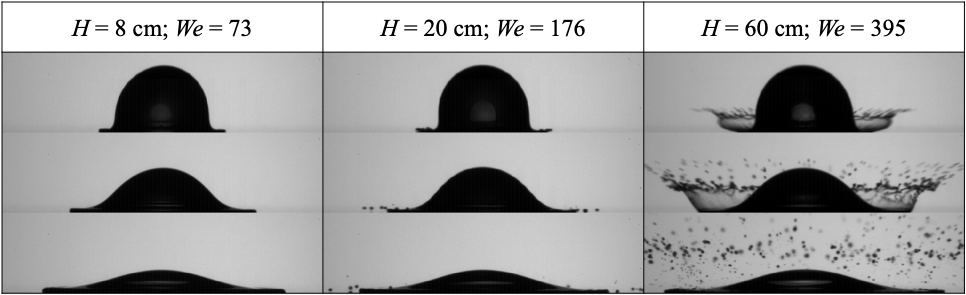}
\caption{\label{fig:processed_img} 
Several examples of the image sequences combined from the images of $z_0/2R_0 = 0.75$, $0.50$, and $0.25$ for (left) a nonsplashing drop with $H = 8~\mathrm{cm}$ and $We = 73$, (middle) a splashing drop near the splashing thresholdwith $H = 20~\mathrm{cm}$ and $We = 176$, and (right) a splashing drop with $H = 60~\mathrm{cm}$ and $We = 395$.
Attributed to specific videographing conditions and digital image processing, the image sequences are highly similar regardless of $H$, $We$, and the outcomes.
}\end{figure}

The focus of the current study is the temporal evolution of the splashing and nonsplashing drops at the early stage of the impact, which has two dynamical regimes, namely pressure impact and self-similar inertial regimes \cite{lagubeau2012spreading}.
Therefore, the videos or image sequences are formed by combining three images: each from the pressure impact regime when the normalized drop central height $z_0/2R_0 = 0.75$, the transition between pressure impact regime and self-similar inertial regime when $z_0/2R_0 = 0.50$, and the self-similar inertial regime when $z_0/2R_0 = 0.25$, respectively.
Here, $z_0$ is central height of the drop (see Fig.~\ref{fig:z0_2R0}) and $z_0/2R_0$ for each dynamical regime is decided based on the study by Lagubeau \textit{et al.} \cite{lagubeau2012spreading}.
The normalized drop central height $z_0/2R_0$ can be understood as the remaining portion of the drop which has yet to impact the solid surface.
For example, when $z_0/2R_0 = 0.75$, three quarters of the drop has yet to impact the solid surface and one quarter of it has already impacted the solid surface.
Note that $z_0/2R_0$ is arranged in descending order as it decreases with impact time.
Attributed to the specific videographing conditions and digital image processing, the image sequences are highly similar regardless of $H$, $We$, and the outcome, as shown in Fig.~\ref{fig:processed_img}.

Fivefold cross validation was performed to ensure the generalizability of the trained feedforward neural networks (FNNs).
For that, the image sequences of each $H$ were segmented into five combinations of training-validation and testing data with the ratio of 80:20.
This is to ensure that the data of each $H$ are included in both training-validation and testing and distributed evenly among the data combinations.
This is reflected in the similar numbers of splashing and nonsplashing data for training-validation or testing among all data combinations, as shown in Table~\ref{tab:data_num}.

\subsection{\label{sec:ann} Feedforward Neural Network (FNN)}

\begin{figure}
\centering
\includegraphics[width=0.75\textwidth]{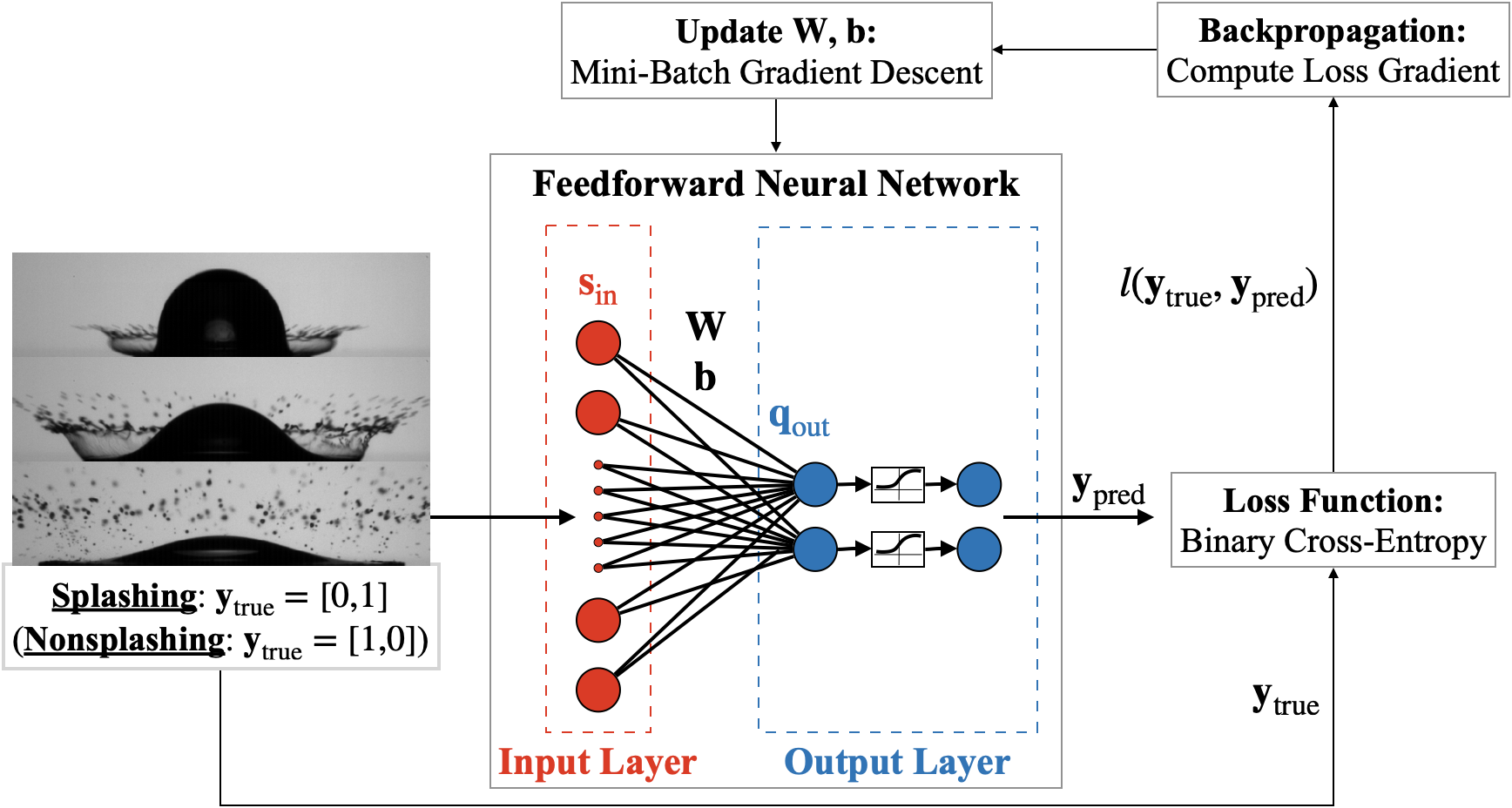}
\caption{\label{fig:fnn_arch_train} Training and architecture of the feedforward neural network (FNN) used to classify image sequences of splashing and nonsplashing drops.}
\end{figure}

Feedforward neural network (FNN) is implemented in the Python programming language on Google Colaboratory \cite{carneiro2018performance} using the libraries of TensorFlow \cite{abadi2016tensorflow}.
Figure~\ref{fig:fnn_arch_train} illustrates the training and the architecture of the FNN used to classify splashing and nonsplashing drops based on the image sequences, which show the temporal evolution of the drops at the early stage of impact.
In the input layer, the input image sequence is flattened into a 1-dimensional column vector $\mathbf{s}_\mathrm{in} \in \mathbb{R}^{M}$, for $M = N_\mathrm{img} h_\mathrm{img} w_\mathrm{img}$, where $N_\mathrm{img}$ is the total number of images in an image sequence, $h_\mathrm{img}$ is the height of an image in pixels, and $w_\mathrm{img}$ is the width of an image in pixels.

Each element of ${\bf s}_\mathrm{in}$ in the input layer (red circles in Fig.~\ref{fig:fnn_arch_train}) is fully connected to each element of ${\bf q}_\mathrm{out}$ in the output layer (blue circles) by a linear function:
\begin{equation}
{\bf q}_\mathrm{out} = {\bf W}{\bf s}_\mathrm{in} + {\bf b}
\label{eq:lin_func_out},
\end{equation}
where ${\bf q}_\mathrm{out} \in \mathbb{R}^{C}$ is the output vector, ${\bf W} \in \mathbb{R}^{C \times M}$ is the weight matrix, and ${\bf b} \in \mathbb{R}^{C}$ is the bias vector.
$C$ is the total number of classes for classification, which are splashing and nonsplashing in this case, and so $C=2$.
The value for each element in ${\bf W}$ and ${\bf b}$, which is initialized using Glorot uniform initializer \cite{glorot2010understanding}, is determined through the training.

In the output layer, each element of ${\bf q}_\mathrm{out}$ is activated by a sigmoid function, which saturates negative values at 0 and positive values at 1, as follows:
\begin{equation}
y_{\mathrm{pred},i} = \sigma(q_{\mathrm{out},i}) = \frac{1}{1+e^{-q_{\mathrm{out},i}}}
\label{eq:sigmoid},
\end{equation}
for $i = 1,\dots,C$, where the activated value ${y}_{\mathrm{pred},i}$ is the element of ${\bf y}_\mathrm{pred} \in \mathbb{R}^{C}$.
${\bf y}_\mathrm{pred} = [y_{\mathrm{pred},1}, y_{\mathrm{pred},2}]$ can be interpreted as a vector that contains the probabilities of an input image sequence to be classified as a nonsplashing drop $y_{\mathrm{pred},1}$ and as a splashing drop $y_{\mathrm{pred},2}$.
For training, ${\bf y}_\mathrm{pred}$ is computed for all train image sequences and compared with the respective true labels ${\bf y}_\mathrm{true}\in \mathbb{R}^{C}$.
The true labels for the image sequences of a splashing drop and a nonsplashing drop are ${\bf y}_\mathrm{true} = [0,1]$ and $[0,1]$, respectively.

Binary cross-entropy loss function is used for the comparison between ${\bf y}_\mathrm{pred}$ and ${\bf y}_\mathrm{true}$ as follows:
\begin{equation}
l({\bf y}_\mathrm{true},{\bf y}_\mathrm{pred})
=\sum_{i=1}^{C}\left[
-{y}_{\mathrm{true},i}\ln({ y}_{\mathrm{pred},i})-(1-{y}_{\mathrm{true},i})\ln(1-{y}_{\mathrm{pred},i})\right]
\label{eq:cross_entropy},
\end{equation}
for $i = 1, \dots, C$, where $l$ is the computed loss.
Owing to this equation, the value of $l$ approaches 0 as ${\bf y}_\mathrm{pred}$ approaches ${\bf y}_\mathrm{true}$ and increases significantly as ${\bf y}_\mathrm{pred}$ varies away from ${\bf y}_\mathrm{true}$.
$l$ is computed during training and validation, but not during testing.

Through the backpropagation algorithm \cite{rumelhart1986learning}, the gradient of $l$ with respect to each element of $\bf W$ and $\bf b$ of the FNN is computed.
The computed gradient determines whether to increase or decrease the value of an element and how much it should be increased or decreased when $\bf W$ and $\bf b$ are updated using the algorithm of mini-batch gradient descent \cite{li2014efficient}.
Regularization of early stopping \cite{prechelt1998early} is applied to determine when to stop updating $\bf W$ and $\bf b$.

The percentage accuracy of the trained FNN is also evaluated using the following equation:
\begin{equation}
\text{accuracy}= \frac{\text{number of correct predictions}}{\text{total number of predictions}} \times 100.
 \end{equation}
The number of correct predictions is determined by the classification threshold.
The trained FNN classifies an image sequence based on the element of ${\bf y}_\mathrm{pred}$ that has a value equal to or greater than that of the classification threshold.
In this study, the classification threshold is fixed at $0.5$.
For example, if the prediction of an image sequence by the trained FNN is ${\bf y}_\mathrm{pred} = [0.25,0.75]$, then the image sequence will be classified as an image of a splashing drop.
Accuracy is computed during training, validation, and testing.
\begin{figure}
\subfloat[]{
\includegraphics[width=0.45\columnwidth]{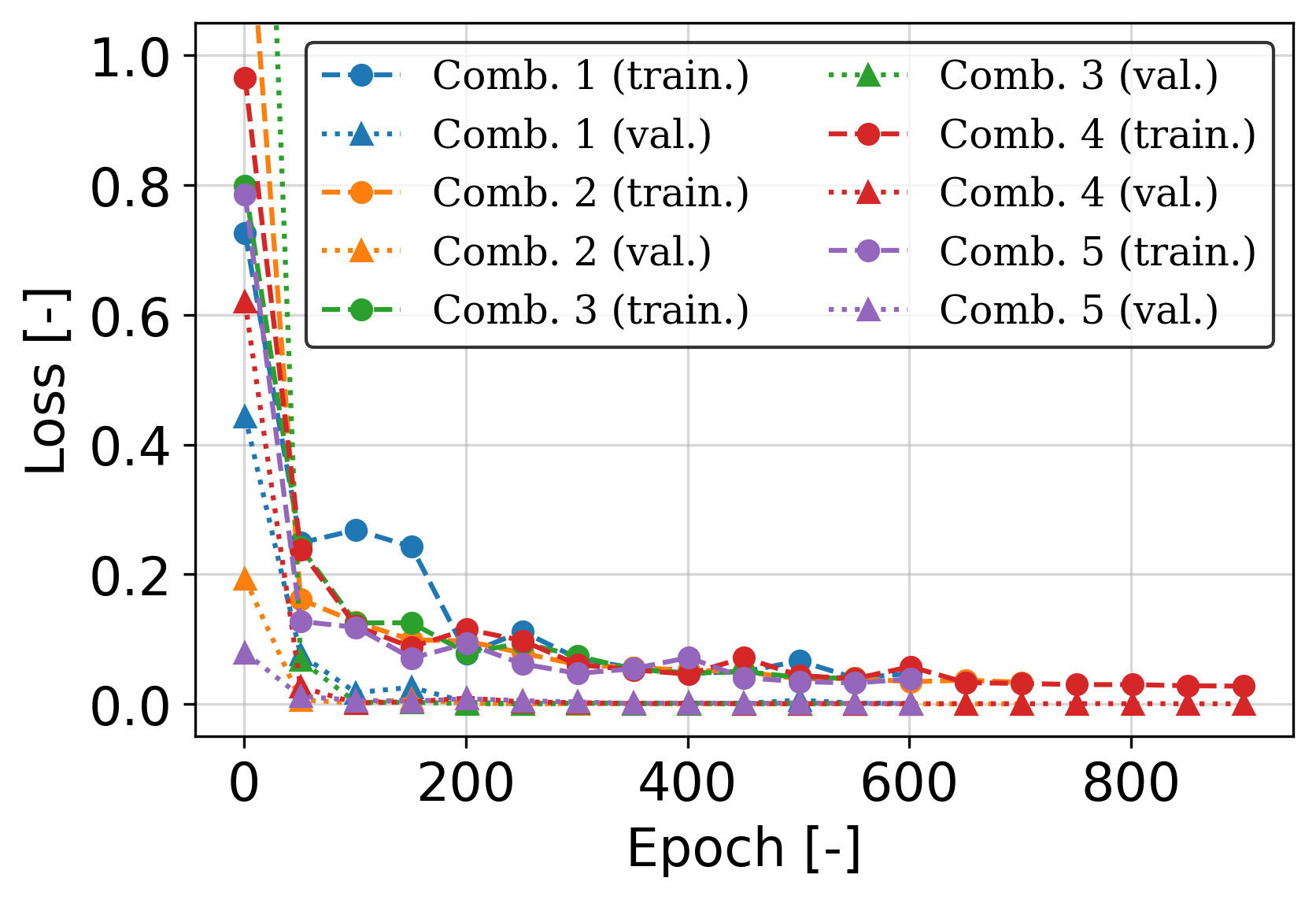}
\label{fig:loss} }
\hfill
\subfloat[]{
\includegraphics[width=0.45\columnwidth]{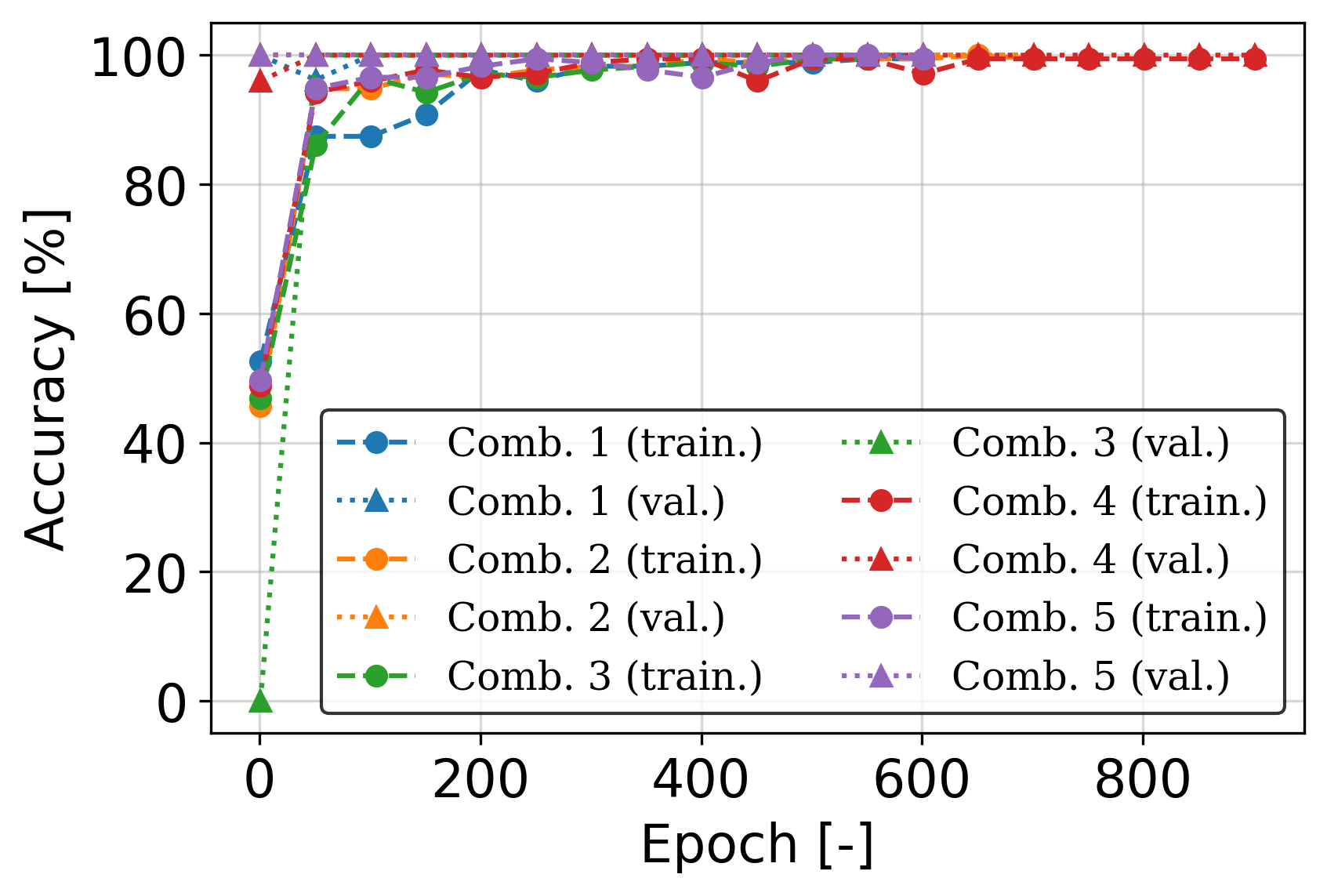}
\label{fig:acc} }
\caption{\label{fig:class_perform} Training and validation of the FNNs for image-sequence classification: (\protect\subref*{fig:loss}) losses and (\protect\subref*{fig:acc}) accuracies after every fifty epochs. 
Losses decrease and approach 0 while accuracies increase and approach 1 as the number of epoch increases.
Comb., combination; train., training loss or accuracy; val. validation loss or accuracy.}
\end{figure}

The training and validation of the FNNs for image-sequence classification are evaluated from the plots of losses and accuracies after every fifty epochs, which are shown in Fig.~\ref{fig:class_perform}.
Here, the number of epochs indicates how many times all training-validation image sequences are fed through the FNN for training.
Losses decrease and approach 0 while accuracies increase and approach 1 as the number of epoch increases.
Early stopping prevented overfitting by stopping the updating of $\bf W$ and $\bf b$ when the losses reach the minimum values.
These trends confirmed that the training and validation were carried out properly and the trained FNNs have achieved desired classification performance.
The trained FNNs are then used to classify test image sequences to check the generalizability.

\section{\label{sec:result}Results and Discussion}
In this section, the results and discussion are presented.
In Section~\ref{sec:class_perform}, the testing of the trained feedforward neural network (FNN) is explained.
As for Section~\ref{sec:fea_ext}, the process for extracting the features used by the FNN to classify splashing and nonsplashing drops is elaborated.
Finally, in Section~\ref{sec:time_evolve}, the physical interpretation and the temporal evolution of the extracted features are discussed.


\subsection{\label{sec:class_perform}Testing of Feedforward Neural Networks (FNNs)}
Testing is the evaluation on the ability of the trained feedforward neural network (FNN) in classifying new image sequences that show the temporal evolution of the drop at the early stage of impact.
The results for all data combinations are shown in Table~\ref{tab:test_result_img_seq}.
Among all combinations, test accuracy in classifying image sequences of splashing and nonsplashing drops is higher than $96\%$.
This indicates that the trained FNNs can identify splashing and nonsplashing drops based on their temporal evolution at the early stage of impact.
Such high accuracy in image-sequence classification of the simple but highly interpretable FNN was made possible with the use the highly similar dataset.
The FNNs can now be analyzed to extract the features that they identify to classify splashing and nonsplashing drops.


\begin{table}
\centering
\caption{\label{tab:test_result_img_seq}
Test accuracy of FNNs trained with different data combinations in classifying image sequences of splashing and nonsplashing drops.
The high test accuracy achieved by the trained FNNs indicates that for highly similar dataset, image-sequence classification can be performed using a simple but highly interpretable FNN.
}
\begin{tabular}{|c|c|c|c|c|c|c|}
\hline
\multirow{2}{*}{Combination}&
\multicolumn{6}{c|}{Test Accuracy} \\\cline{2-7} &
\multicolumn{2}{c|}{Splashing} & \multicolumn{2}{c|}{Nonsplashing} & \multicolumn{2}{c|}{Total} \\ \hline
1& 26/27& 96\%& 21/21& 100\%& 47/48& 98\% \\ \hline
2& 28/29& 97\%& 22/22& 100\%& 50/51& 98\% \\ \hline
3& 26/28& 93\%& 23/23& 100\%& 49/51& 96\% \\ \hline
4& 27/27& 100\%& 23/23& 100\%& 50/50& 100\% \\ \hline
5& 28/30& 93\%& 19/19& 100\%& 47/49& 96\%
\\ \hline
\end{tabular}
\end{table}

\subsection{\label{sec:fea_ext} Extraction of Features of Splashing and Nonsplashing Drops}
To extract the features which the trained feedforward neural networks (FNNs) identify for classification, the important pixel positions were determined by reshaping and visualizing the trained weight matrix ${\bf W}$ as follows.
The matrix form of ${\mathbf W} \in \mathbb{R}^{C \times M}$ is
\begin{equation}
{\mathbf W} = 
\begin{bmatrix}
W_{1,1} &W_{1,2} &\dots &W_{1,M}\\
W_{2,1} &W_{2,2} &\dots &W_{2,M}\\
\end{bmatrix}
=
\begin{bmatrix}
{\mathbf w}_{1}\\
{\mathbf w}_{2}\\
\end{bmatrix}
\label{eq:W_h},
\end{equation}
where the elements ${\mathbf w}_{1}$ and ${\mathbf w}_{2}$ are the vector elements of the weight matrix for computing the probability of an image sequence being nonsplashing and splashing drops, respectively.
For visualization, ${\mathbf w}_{2} \in \mathbb{R}^M$, was divided into three smaller vectors with a size of $h_\mathrm{img} w_\mathrm{img}$.
Each of these smaller vectors corresponds to the images of the dynamical regimes of pressure impact ($z_0/2R_0 = 0.75$), transition ($z_0/2R_0 = 0.50$), and self-similar inertial ($z_0/2R_0 = 0.25$), respectively, in the image sequences.
The vectors were then reshaped in row-major order into a two-dimensional matrix of $h_\mathrm{img} \times w_\mathrm{img}$, which is the shape of the images in the input image sequence, and viewed as colormaps.
\begin{figure}
\centering
\includegraphics[width=0.7\textwidth]{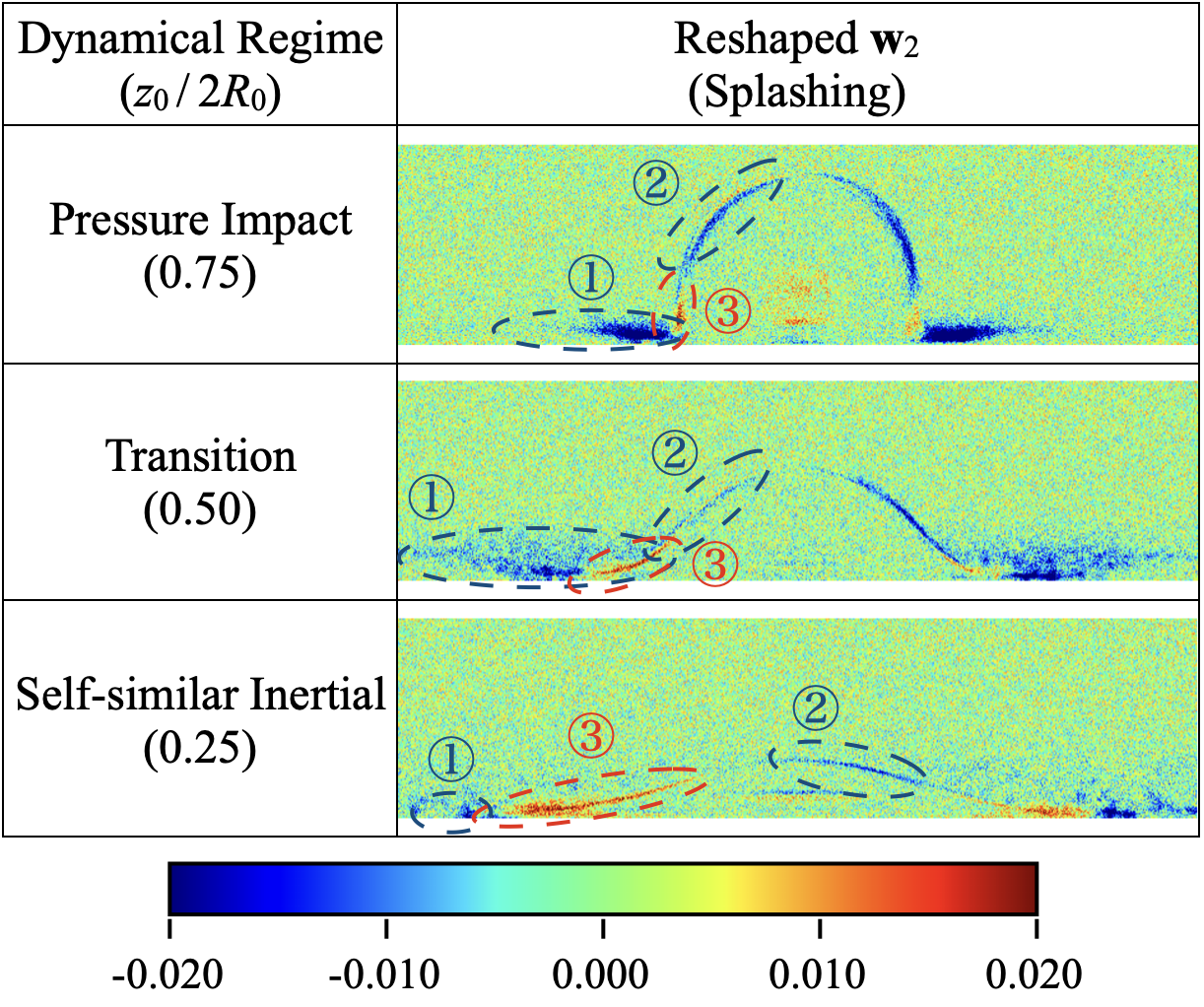}
\caption{
\label{fig:trained_w} 
Colormaps of the reshaped vectors of ${\mathbf w}_{2}$, which correspond to the images of the dynamical regimes of pressure impact ($z_0/2R_0 = 0.75$), transition ($z_0/2R_0 = 0.50$), and self-similar inertial ($z_0/2R_0 = 0.75$), respectively, of the FNN trained with combination 1.
\textcircled{1}, \textcircled{2}, and \textcircled{3} indicate the distributions of values with large magnitudes, which correspond to the features of typical splashing and nonsplashing drops indicated by \textcircled{1}, \textcircled{2}, and \textcircled{3} in Fig.~\ref{fig:typ_drop}.
The distributions are similar for the FNNs trained other combinations.
}
\end{figure}
\begin{figure}
\centering
\includegraphics[width=1.0\textwidth]{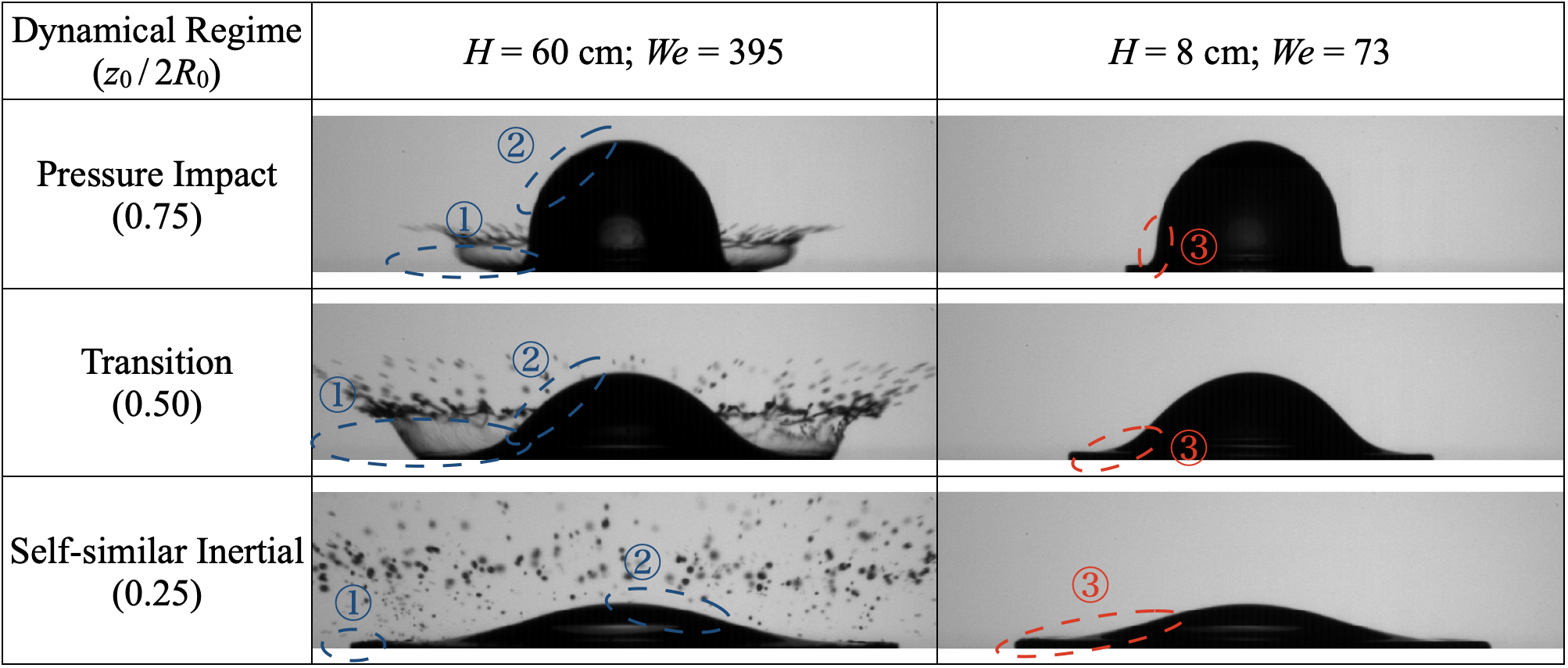}
\caption{
\label{fig:typ_drop} 
Image sequences of typical splashing ($H = 60~\mathrm{cm}$ and $We = 395$ of Fig.~\ref{fig:processed_img}) and nonsplashing drops ($H = 8~\mathrm{cm}$ and $We = 73$ of Fig.~\ref{fig:processed_img}).
\textcircled{1} indicates the ejected secondary droplets around the aerodynamically lifted lamella of the splashing drop, \textcircled{2} the relatively high contour of the main body of the splashing drop, and \textcircled{3} the relatively short and thick lamella of the nonsplashing drop.
These areas correspond to the distribution of values with large magnitudes in the colormaps of the reshaped vectors of ${\mathbf w}_2$ shown in Fig.~\ref{fig:trained_w}.
}
\end{figure}

For explanation, the colormaps of the reshaped vectors of ${\mathbf w}_{2}$ for the FNN trained with combination 1 is presented in Fig.~\ref{fig:trained_w}.
The blue-green-red (BGR) scale is from -0.020 to 0.020.
Note that only those with combination 1 are shown because they are similar to those with other combinations.
In the colormaps, the distribution of the extreme values, i.e., values of large magnitudes, show the important features which help the FNN to classify splashing and nonsplashing drops, where the extreme negative values shown in blue are the features of splashing drops, while the extreme positive values shown in red are the features of nonsplashing drops.
The process of how the extreme negative and positive values of the trained weight $\bf W$ help the FNN in image-sequence classification is similar to that in the image classification reported in our previous study \cite{yee2022image}.
Therefore, the explanation is omitted in this paper.
Moreover, the colormaps of the reshaped vectors of ${\mathbf w}_1$, which computes the nonsplashing probability of an image sequence, and the related discussion are not included in this paper because the extracted features and the way they help the FNN in image-sequence classification are the same as those of ${\mathbf w}_2$.

By comparing the distribution of the extreme values in the colormaps of the reshaped vectors of ${\mathbf w}_2$ with the image sequences of a typical splashing drop and a typical nonsplashing drop in Fig.~\ref{fig:typ_drop}, it is found that the distribution of the values of large magnitudes resembles the temporal evolution of the impacting drops.
Remarkably, the extreme negative values (blue) are distributed around the areas that correspond to the temporal evolution of \textcircled{1} the ejected secondary droplets around the aerodynamically lifted lamella\footnote{Lamella is the thin liquid sheet ejected from the side of an impacting drop.} and \textcircled{2} the relatively high contour of the main body of a splashing drop.
On the other hand, the extreme positive values (red) are found around the area that correspond to the temporal evolution of \textcircled{3} the relatively short and thick lamella of a nonsplashing drop.
The distribution of these extreme values in each reshaped vector is symmetric, except for the distribution of the extreme negative values at the area around \textcircled{2} of the reshaped vector that corresponds to $z_0/2R_0 = 0.25$.
These findings indicate that the trained FNN identifies \textcircled{1} the ejected secondary droplets around the aerodynamically lifted lamella and \textcircled{2} the relatively high contour of the main body as the features of a splashing drop, while \textcircled{3} the relatively short and thick lamella as the feature of a nonsplashing drop.
Such features are similar to the instantaneous image features extracted through image classification performed in our previous study \cite{yee2022image}.
The physical interpretation and the temporal evolution of these features are discussed in Section~\ref{sec:time_evolve}.


For the trained bias $\bf b$, the order of magnitude of the elements is $10^{-3}$, which is much smaller than the elements of ${\bf q}_\mathrm{out}$ computed by the trained FNNs.
On the other hand, among the test image sequences of all data combinations, the smallest absolute value of ${q}_\mathrm{out}$ is 0.43.
Therefore, ${\bf q}_{out} \approx {\bf W}{\bf s}_{in}$.
This indicates that the trained $\bf b$ did not affect the classification of the FNN and is negligible.

\subsection{\label{sec:time_evolve}Physical Interpretation and Temporal Evolution of Features of Splashing and Nonsplashing Drops}
\begin{figure}
\centering
\includegraphics[width=0.95\textwidth]{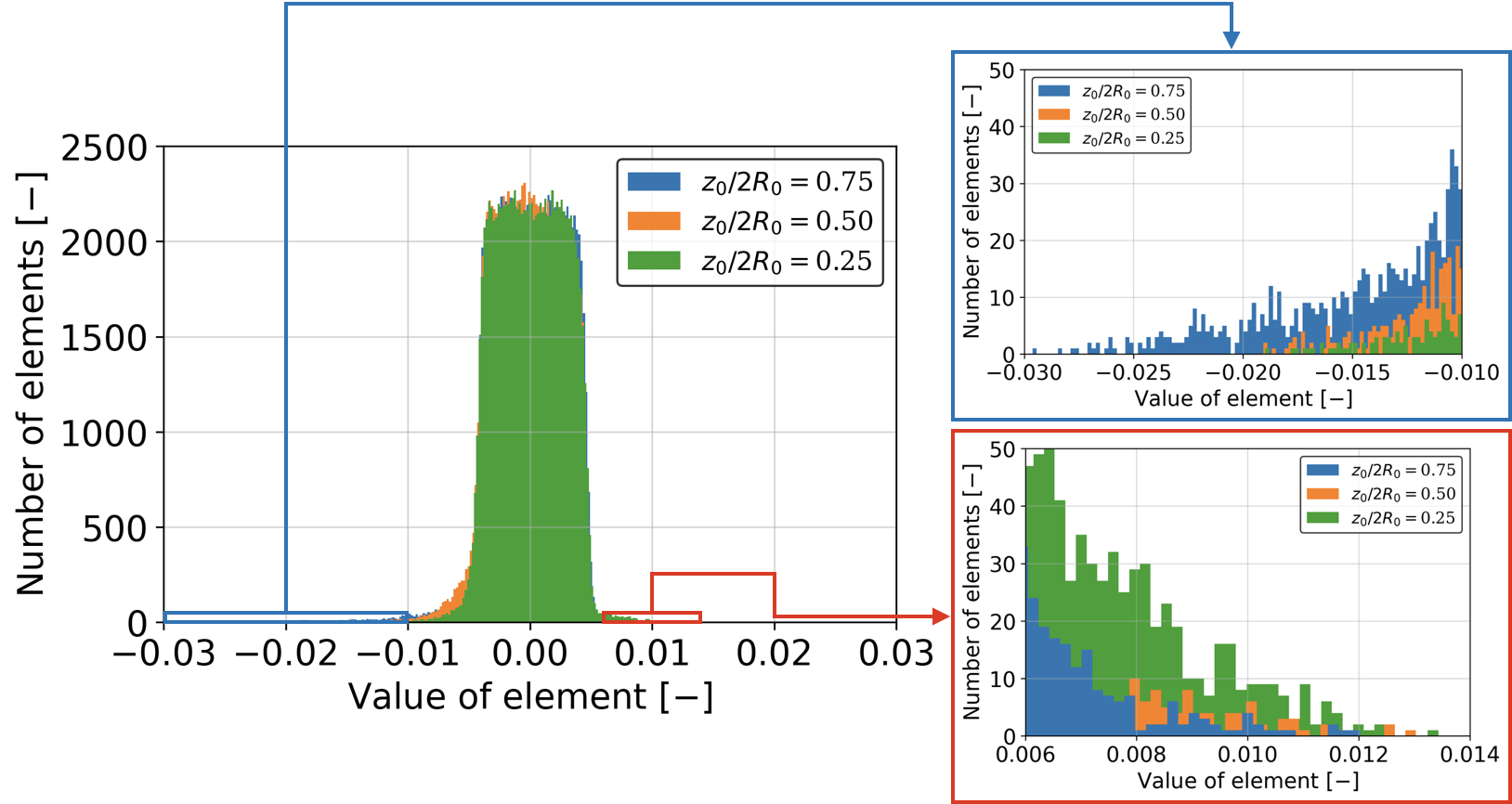}
\caption{
\label{fig:hist} 
Number of elements for each value in each reshaped vector of ${\mathbf w}_{2}$.
For the analysis on the features of splashing drops, the upper inset shows the number of elements with extreme negative values, which are less than $-0.010$.
The number of elements with extreme negative values is highest when $z_0/2R_0 = 0.75$, followed by $z_0/2R_0 = 0.50$, and is lowest when $z_0/2R_0 = 0.25$.
On the other hand, for the analysis on the features of nonsplashing drops, the lower inset shows the number of elements with extreme positive values, which are greater than $0.006$.
For the elements with values $\geq 0.008$, their number increases from $z_0/2R_0 = 0.75$ to $0.50$ and has the highest number when $z_0/2R_0 = 0.25$. 
}
\end{figure}
In this subsection, the physical interpretation and the temporal evolution of the features of the splashing and nonsplashing drops are analyzed and discussed.

Careful observation on the colormaps of the reshaped vectors of ${\mathbf w}_2$ in Fig.~\ref{fig:trained_w} shows that the distribution of extreme negative values decreases while that of extreme positive values increases in the order of $z_0/2R_0 = 0.75$, 0.05, and 0.25.
Such observation is confirmed by the histogram for the elements in each reshaped vector of ${\mathbf w}_2$, which is plotted using the hist() function of Matplotlib library \cite{hunter2007matplotlib} in Fig.~\ref{fig:hist}.
Note that most of the elements have the values around 0.00 and do not affect the classification of the trained FNN.
They occupy areas that correspond to the image background and the main body of the impacting drop, which are highly similar for both splashing and nonsplashing drops.
Therefore, the important elements which affect the classification are those with extreme values, where the extreme negative and positive values indicate the features of splashing and nonsplashing drops, respectively.
For the analysis on the features of splashing drops, the upper inset shows the number of elements with extreme negative values, which are less than $-0.010$.
The number of elements with extreme negative values is highest when $z_0/2R_0 = 0.75$, followed by $z_0/2R_0 = 0.50$, and is lowest when $z_0/2R_0 = 0.25$.
On the other hand, for the analysis on the features of nonsplashing drops, the lower inset shows the number of elements with extreme positive values, which are greater than $0.006$.
For the elements with values $\geq 0.008$, their number increases from $z_0/2R_0 = 0.75$ to $0.50$ and has the highest number when $z_0/2R_0 = 0.25$.
As $z_0/2R_0$ decreases with impact time, the findings here indicate that as impact time elapses, the features of splashing and nonsplashing drops become less and more important, respectively, in helping the trained FNN in classification.

\begin{figure}
\centering
\includegraphics[width=0.6\textwidth]{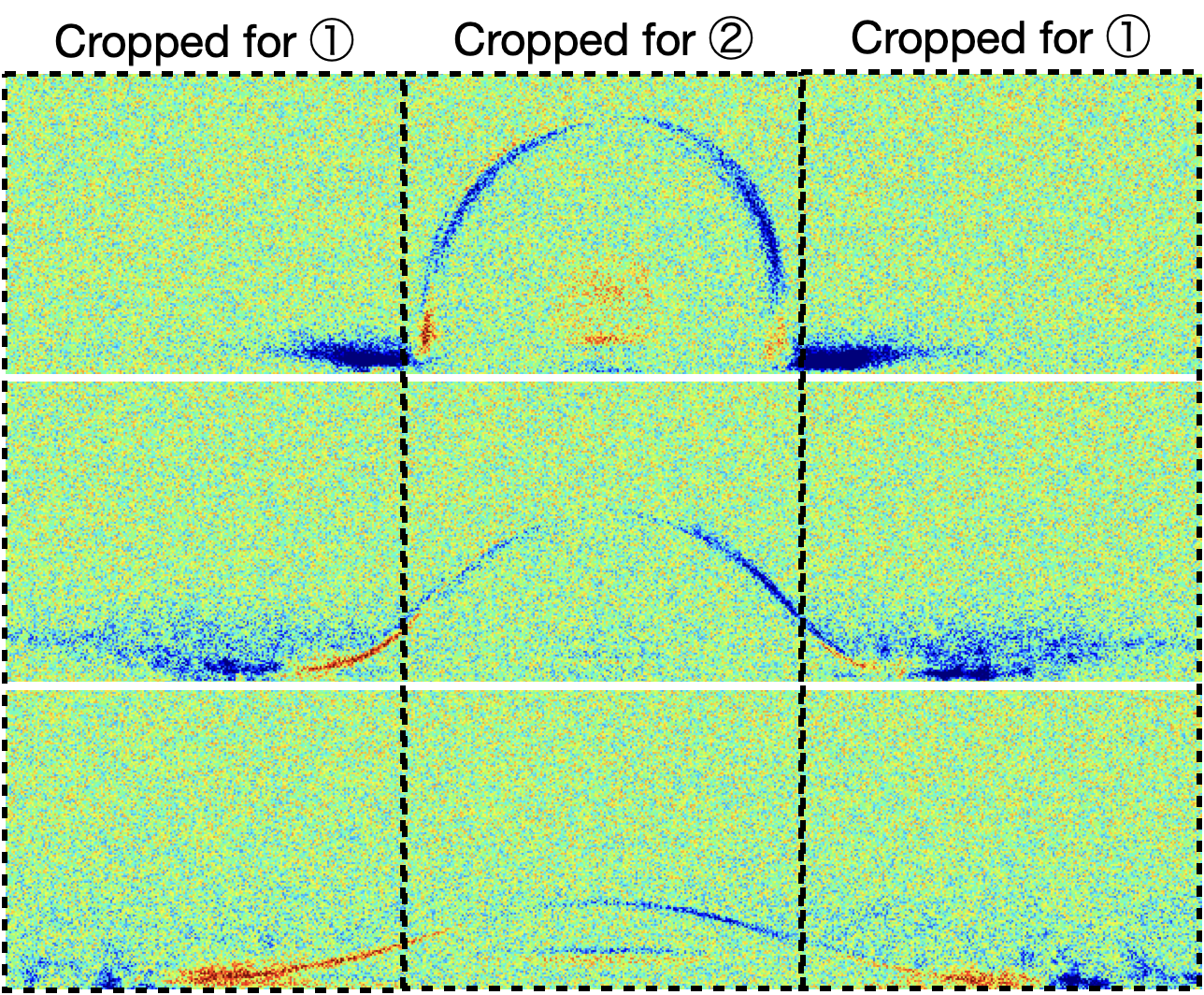}
\caption{
\label{fig:trained_w_crop} 
Colormaps of the reshaped vectors of ${\mathbf w}_{2}$ are cropped horizontally into three equal parts, where the left and right parts are cropped for plotting the histogram for \textcircled{1} (Fig.~\ref{fig:hist_lam}) while the center part is cropped for plotting the histogram for \textcircled{2} (Fig.~\ref{fig:hist_mb}).
}
\end{figure}

\begin{figure}
\subfloat[]{
\includegraphics[width=0.45\columnwidth]{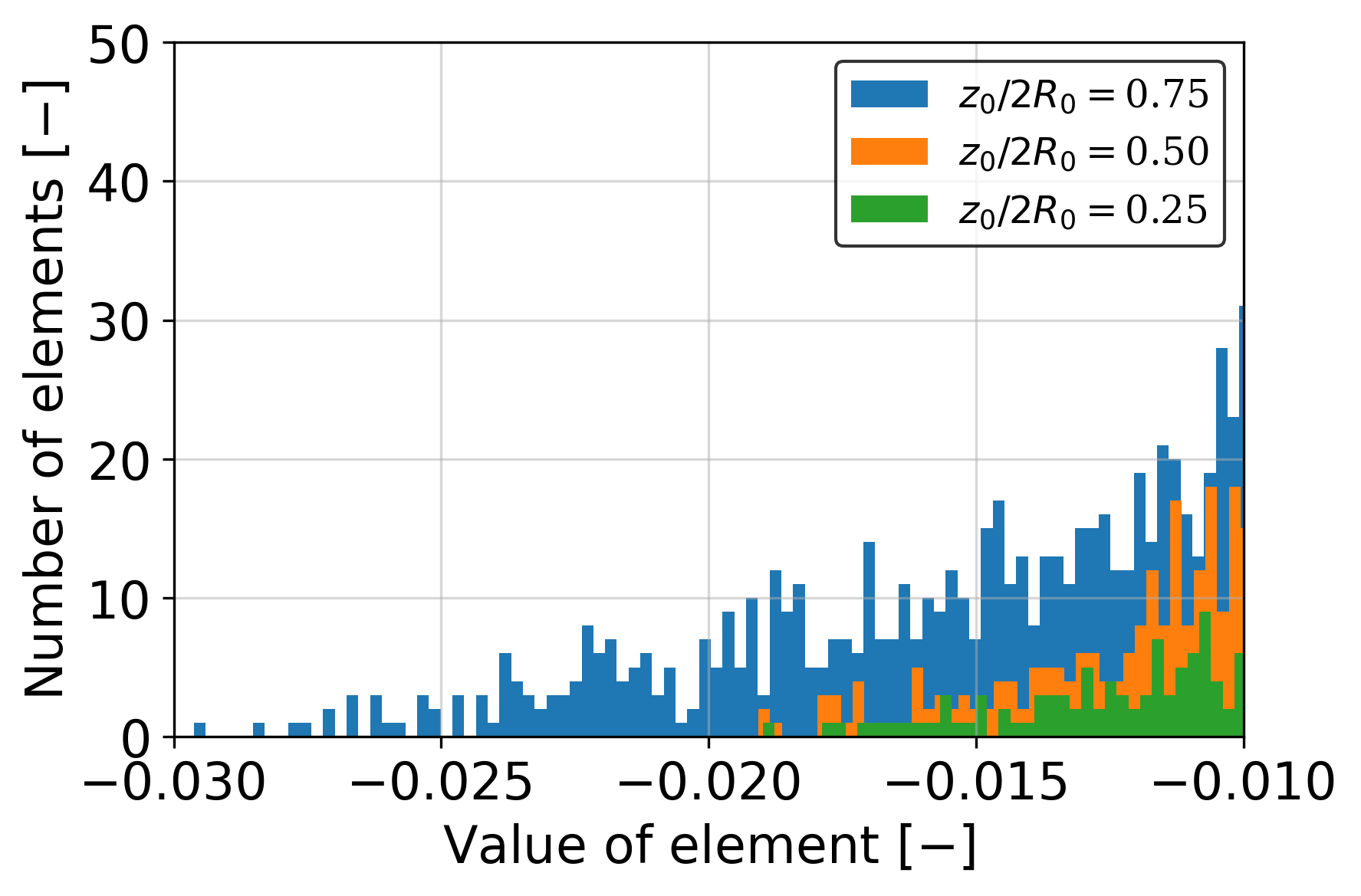}
\label{fig:hist_lam} }
\hfill
\subfloat[]{
\includegraphics[width=0.45\columnwidth]{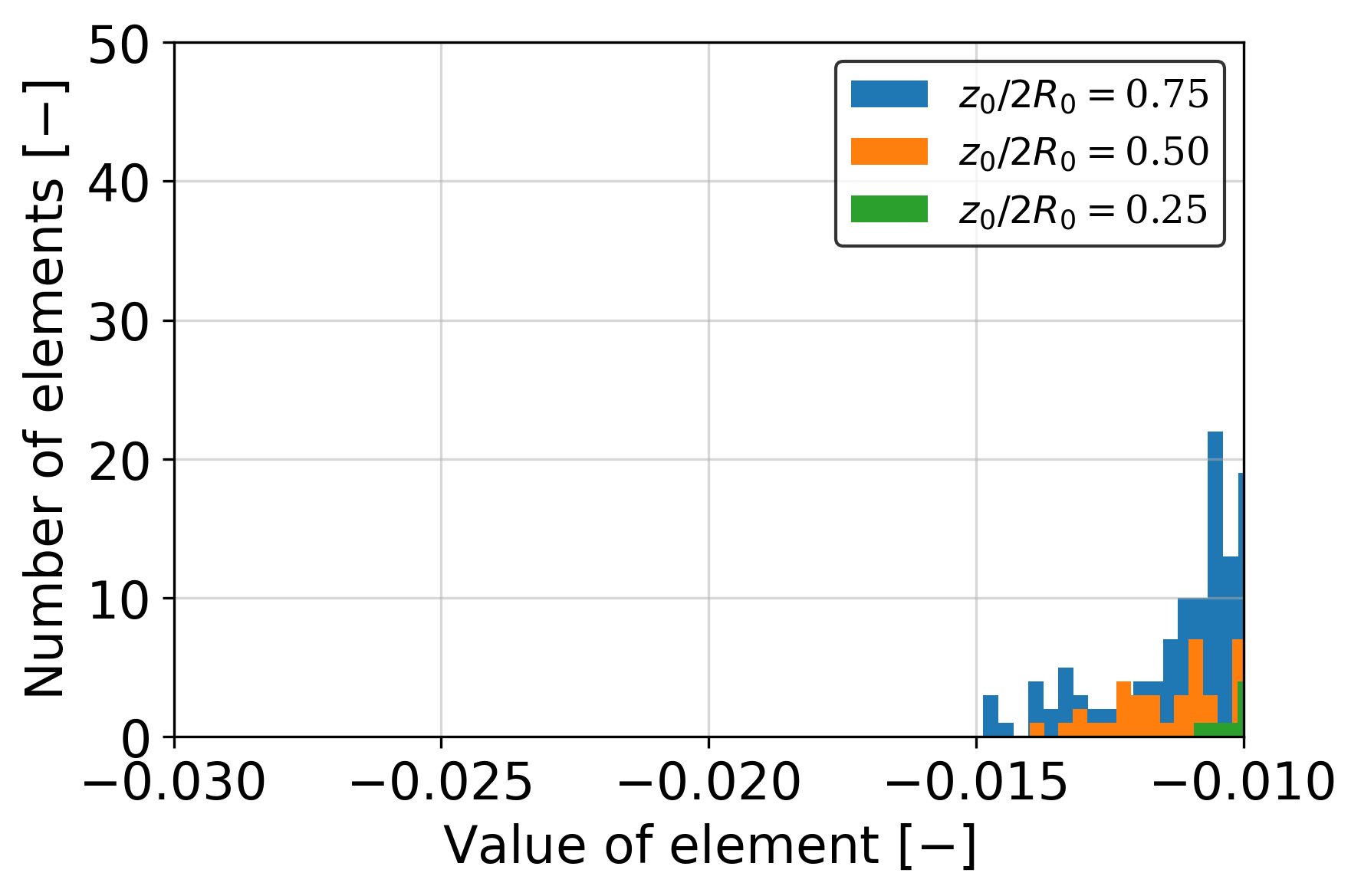}
\label{fig:hist_mb} }
\caption{\label{fig:hist_neg_all} Number of elements with extreme negative values, which are less than $-0.010$, in (\protect\subref*{fig:hist_lam}) the left and right parts (histogram for \textcircled{1}) and (\protect\subref*{fig:hist_mb}) the center (histogram for \textcircled{2}) in each reshaped vectors of ${\mathbf w}_{2}$.
In (\protect\subref*{fig:hist_lam}), there are more elements with extreme negative values and the minimum value is much lower, as compared to (\protect\subref*{fig:hist_mb}).
This indicates that \textcircled{1} is a more dominant feature of a splashing drop.
In terms of temporal evolution, for both (\protect\subref*{fig:hist_lam}) and (\protect\subref*{fig:hist_lam}), the number of elements with extreme negative values is the highest when $z_0/2R_0 = 0.75$ and decreases together with $z_0/2R_0$.
}
\end{figure}

The analysis of the splashing features shown in the upper inset of Fig.~\ref{fig:hist} correspond to the temporal evolution of both \textcircled{1} the ejected secondary droplets around the aerodynamically lifted lamella and \textcircled{2} the relatively high contour of the main body.
However, it is necessary to analyze each of the features individually.
For that, a histogram is plotted for each of the feature by cropping the reshaped vectors of ${\mathbf w}_{2}$ into three equal parts as shown in Fig.~\ref{fig:trained_w_crop}, where the left and right areas are cropped for plotting the histogram for \textcircled{1} and the middle area is cropped for plotting the histogram for \textcircled{2}.
The histograms are shown in Fig.~\ref{fig:hist_neg_all}.
By comparing both histograms, there are more elements with extreme negative values in the areas cropped \textcircled{1} with a much lower minimum value, indicating that \textcircled{1} is a more dominant feature of a splashing drop from the perspective of the trained FNN.
This is as expected because the areas cropped for \textcircled{1} are two times larger than that for \textcircled{2}.
In addition, \textcircled{1} represents the typical definition of a splashing drop, namely, the ejection of secondary droplets from the impacting drop \cite{josserand2016drop, yarin2006drop}.
However, it is necessary to point out that the extreme negative values in the area cropped for \textcircled{1} only occupy the lower half of each of the reshaped vectors of ${\mathbf w}_2$.
This indicates that the trained FNN only checks the ejected secondary droplets around the aerodynamically lifted lamella.
This can be related to the physical theory introduced by Riboux and Gordillo \cite{riboux2014experiments} that attributes splashing to the aerodynamic lift force that acts on the lamella of the impacting drop.

In terms of temporal evolution, for both \textcircled{1} and \textcircled{2}, the number of elements with extreme negative values is the highest when $z_0/2R_0 = 0.75$ and decreases together with $z_0/2R_0$.
For \textcircled{1}, this is because when $z_0/2R_0 = 0.75$, the ejected secondary droplets are more concentrated around the aerodynamically lifted lamella.
As impact time elapses, the ejected secondary droplets travel further and become more scattered.
As for \textcircled{2}, to the best of the authors' knowledge, it is first identified as a feature of a splashing drop in our previous study \cite{yee2022image}.
Since the physical interpretation is still unclear, it is difficult to discuss why the feature becomes less prevalent as time elapses.
Eventually, the distribution of the extreme negative values that correspond to \textcircled{2} become asymmetric and only appear on the right half of the colormap that corresponds to $z_0/2R_0 = 0.25$.

As for the features of a nonsplashing drop that are indicated by the extreme positive values, they correspond to the temporal evolution of \textcircled{3} the relatively short and thick lamella.
Owing to the lower Weber number $We$ of a nonsplashing drop, the lamella of a nonsplashing drop ejects at a later time and a lower velocity as compared to a splashing drop. \cite{riboux2014experiments,riboux2017boundary}.
Therefore, when $z_0/2R_0 = 0.75$, the lamella of a nonsplashing drop just starts to be ejected.
As time elapses, although the lamella of a nonsplashing drop grows, it is still much shorter as compared to the fast-growing lamella of a splashing drop.
Nevertheless, the lamella of a nonsplashing drop grows thicker thus easier to be identified by the FNN.
This explains why the vector corresponds to $z_0/2R_0 = 0.75$ has the highest number of elements with extreme positive values.

To summarize this subsection, the trained FNN captured the important features that distinguish splashing and nonsplashing drops from every image in the image sequences, which show the temporal evolution of the drop during the early stage of impact.
Moreover, it also captured the temporal evolution of these features.
The physical interpretation of these features and the respective temporal evolution have been identified except for \textcircled{2} the difference in contour height of the main body between splashing and nonsplashing drops.



\section{\label{sec:conclusion} Conclusion and Outlook}
In this study, classification of splashing and nonsplashing drops has been successfully performed based on the image sequences that show the temporal evolution of the drops during the early stage of impact.
Test accuracy higher than 96\% has been achieved using a highly interpretable feedforward neural network (FNN) with zero hidden layer.
This was made possible by using the highly similar image sequences for training-validation and testing.

The features used by trained FNN to identify a splashing drop is the temporal evolution of the ejected secondary droplets around the aerodynamically lifted lamella and the relatively high contour of the main body.
Whereas to identify a nonsplashing drop, the FNN checks the temporal evolution of the relatively short and thick lamella.
These features are similar to those extracted through the image-based classification reported in our previous study \cite{yee2022image}.
The temporal evolution of these features were analyzed.
The analysis shows that from the perspective of the trained FNNs, the splashing features become less prevalent while the nonsplashing features become more prevalent as time elapses.

Eventually, we hope that our study can be developed into a data-driven simulation for modeling the deformation of a splashing drop during the impact on a solid surface.
Therefore, for the next step of this study, it is suggested to add more images to the image sequences so that more details of the temporal evolution of the drop during the impact can be captured by the artificial neural network (ANN).
Moreover, it is also recommended to train the ANN to perform next-frame prediction for predicting the temporal evolution of the drop during the impact.

\section*{Acknowledgments}
This work was funded by Epson International Scholarship Foundation, Rotary Yoneyama Memorial Foundation, Japan Society for the Promotion of Science (Grant Nos. 20H00223, 20H00222, and 20K20972) and Japan Science and Technology Agency PRESTO (Grant No. JPMJPR21O5). 
The authors would also like to thank Dr. Masaharu Kameda (Professor, Tokyo University of Agriculture and Technology) for his valuable discussions and suggestions.

\bibliography{sample}

\end{document}